\documentstyle[11pt,paspconf,psfig]{article}

\begin{document}

\title{Setting the stellar evolution clock for intermediate age populations} 
\author{Raul Jimenez} 
\affil{Raul Jimenez, Institute for Astronomy, University of Edinburgh, 
Blackford Hill, Edinburgh EH9 3HJ, UK}

\begin{abstract}
In this invited talk I show how the {\it reddest and rarest} galaxies at high
redshift ($z \simeq 1.5$) can be used to set the stellar evolution clock. I
argue that one can confidently compute the collapse redshift of these
objects. This yields to a high collapse redshift ($z>6$) and therefore their
age is well constrained (in all cosmologies) between 3 and 4 Gyr. I also show
that this is, indeed, the age derived using a variety of synthetic stellar
population models when proper statistical tools are used to analyse their
observed spectral energy distribution. This allows me to conclude that {\it
all} stellar population models yield to the same consistent age for these
galaxies, i.e. about 3.5 Gyr and that the stellar clock is properly set. Low
ages are therefore excluded with high confidence.

\end{abstract}

\keywords{Cosmology, synthetic stellar populations, stellar evolution}

\section{Introduction}

Traditionally, the stellar clock has been set on stellar objects at $z=0$,
namely the Sun. Here I describe how one can find passively evolving galaxies
at high-redshift and use them to set the stellar clock of stellar populations
of intermediate age.

The most problematic issue is how to find galaxies at high redshift. The study
of `normal' star-forming galaxies at $z > 2$ has developed into a booming
astronomical industry over the past 4 years (e.g. Steidel et al. 1998). Since
most high-redshift galaxies are optically selected, one is biased towards blue
objects that show recent star formation, i.e.  biased towards composite
stellar populations with several episodes of star formation. In this way, one
can only hope to determine the age of the {\it youngest} stars at
high-redshift, not a very useful age indicator.  Nevertheless, a few valiant
efforts have been carried out in order to determine the age of the stellar
populations in high-redshift galaxies (Chambers \& Charlot 1990).

A more promising way to find passively evolving objects is utilizing radio
galaxies. One of the cleanest results in extra-galactic astronomy is that all
powerful ($P > 10^{24}$ WHz$^{-1}$sr$^{-1}$) radio sources in the present-day
universe are hosted by giant ellipticals. It is then reasonable to assume that
high-redshift radio sources also reside in ellipticals or their
progenitors. By selecting radio galaxies at mJy flux levels we (Dunlop et
al. 1996, Spinrad et al. 1997, Dunlop 1998, Dey et al. 1999) have shown that
that it is possible to find examples of well evolved galaxies at $z \sim 1.5$
whose near-ultraviolet spectrum is uncontaminated by a recent burst of star
formation. Keck spectroscopy of these objects has yielded the first detection
of stellar absorption features from {\it old} stars at $z \leq 1.5$ and thus
the first {\it reliable} age-dating of high-redshift objects. The best example
in our sample is 53W069. Using WPFC2 and NICMOS images below and above the
4000 \AA \, break we have verified that the above also holds for its
morphological properties and scalelengths (Dunlop 1998). Using a 2-dimensional
fitting code it is possible to show that 53W069 is consistent with a $r^{1/4}$
law and inconsistent with a exponential disc profile. Furthermore, a physical
half-light radius of $r_{e} \approx 4$ kpc has been obtained assuming
$\Omega=1$ and $H_0 = 50$ km s$^{-1}$ Mpc$^{-1}$, which lies exactly in the
Kormendy relation for ellipticals (Dunlop 1998).

The traditional approach would be to use synthetic stellar population models
to derive the age of 53W069 from its observed spectral energy distribution.
Here I argue, that it is possible to use observations of the abundances and
clustering of high-redshift galaxies to estimate the power spectrum on small
scales, and thus constrain the age of 53W069 without the need of stellar
evolution physics knowledge. The following section summarizes the results of
this exercise, as given by Peacock et al. (1998).

\section{Small-scale power spectrum}

\subsection{Press-Schechter apparatus}

The standard framework for interpreting the abundances of high-redshift
objects in terms of structure-formation models, was outlined by Efstathiou \&
Rees (1988).  The formalism of Press \& Schechter (1974) gives a way of
calculating the fraction $F_c$ of the mass in the universe which has collapsed
into objects more massive than some limit $M$:
\begin{equation}
 F_c(>M,z) = 
 1 - {\rm erf}
\,\left[ {\delta_c \over \sqrt{2}\, \sigma(M)}\right].
\end{equation}
Here, $\sigma(M)$ is the rms fractional density contrast obtained by filtering
the linear-theory density field on the required scale. In practice, this
filtering is usually performed with a spherical `top hat' filter of radius
$R$, with a corresponding mass of $4 \pi \rho_b R^3/3 $, where $\rho_b$ is the
background density.  The number $\delta_c$ is the linear-theory critical
overdensity, which for a `top-hat' overdensity undergoing spherical collapse
is $1.686$ -- virtually independent of $\Omega$. This form describes numerical
simulations very well (see e.g.  Ma \& Bertschinger 1994).  The main
assumption is that the density field obeys Gaussian statistics, which is true
in most inflationary models.  Given some estimate of $F_c$, the number
$\sigma(R)$ can then be inferred. Note that for rare objects this is a
pleasingly robust process: a large error in $F_c$ will give only a small error
in $\sigma(R)$, because the abundance is exponentially sensitive to $\sigma$.

Total masses are of course ill-defined, and a better quantity to use is the
velocity dispersion.  Virial equilibrium for a halo of mass $M$ and proper
radius $r$ demands a circular orbital velocity of $V_c^2 = {GM \over r}$.  For
a spherically collapsed object this velocity can be converted directly into a
Lagrangian comoving radius which contains the mass of the object within the
virialization radius (e.g. White, Efstathiou \& Frenk 1993):
\begin{equation}
R / {\,h^{-1}\,{\rm Mpc}}= {2^{1/2}[V_c/100 {\;{\rm km\,s^{-1}}}] \over
\Omega_m^{1/2}(1+z_c)^{1/2}f_c^{1/6}}.
\end{equation} 
Here, $z_c$ is the redshift of virialization; $\Omega_m$ is the {\it
present\/} value of the matter density parameter; $f_c$ is the density
contrast at virialization of the newly-collapsed object relative to the
background, which is adequately approximated by $f_c=178/\Omega_m^{0.6}(z_c)$,
with only a slight sensitivity to whether $\Lambda$ is non-zero (Eke, Cole \&
Frenk 1996). For isothermal-sphere haloes, the velocity dispersion is
$\sigma_v=V_c/\sqrt{2}$.  Given a formation redshift of interest, and a
velocity dispersion, there is then a direct route to the Lagrangian radius
from which the proto-object collapsed.

\subsection{Abundances and masses of high-redshift objects}

Three classes of high-redshift object can be used to set constraints on the
small-scale power spectrum at high redshift:

\noindent
{\bf (1) Damped Lyman-$\alpha$ absorbers}

If the fraction of baryons in the virialized dark matter halos equals the
global value $\Omega_{\ss B}$, then data on these systems can be used to infer
the total fraction of matter that has collapsed into bound structures at high
redshifts (see Peacock et al 1998 and refs. therein). Therefore, 
\begin{equation}
F_c = {\Omega_{\ss HI}\over \Omega_{\ss B}} \simeq 0.12h
\end{equation}
for these systems.  In this case alone, an explicit value of $h$ is required
in order to obtain the collapsed fraction; $h=0.65$ is assumed and we have
adopted $\Omega_{\ss B}h^2=0.02$.

\noindent
{\bf (2) Lyman-limit galaxies}

Steidel et al. (1996) identified star-forming galaxies between $z=3$ and 3.5
by looking for objects with a spectral break redwards of the $U$ band. Steidel
et al. give the comoving density of their galaxies as
\begin{equation}
N(\Omega=1) \simeq 10^{-2.54} \; ({\,h^{-1}\,{\rm Mpc}})^{-3}.
\end{equation}

Direct dynamical determinations of these masses are still lacking in most
cases. Steidel et al. attempt to infer a velocity width by looking at the
equivalent width of the C and Si absorption lines. These are saturated lines,
and so the equivalent width is sensitive to the velocity dispersion; values in
the range
\begin{equation}
\sigma_v\simeq 180 - 320 {\;{\rm km\,s^{-1}}}
\end{equation}
are implied. In practice, this uncertainty in the velocity does not produce an
important uncertainty in the conclusions.

\noindent
{\bf (3) Red radio galaxies}

This is the set of observations for which we wish to determine their collapse
redshift. Two extremely red galaxies were found at $z=1.43$ and $1.55$, over an
area $1.68\times 10^{-3}\; \rm sr$, so a minimal comoving density is from one
galaxy in this redshift range:
\begin{equation}
N(\Omega=1) 10^{-5.87} \; ({\,h^{-1}\,{\rm Mpc}})^{-3}.
\end{equation}
This figure is comparable to the density of the richest Abell clusters, and is
thus in reasonable agreement with the discovery that rich high-redshift
clusters appear to contain radio-quiet examples of similarly red galaxies
(Dickinson 1995).

Since the velocity dispersions of these galaxies are not observed, they must
be inferred indirectly. This is possible because of the known present-day
Faber-Jackson relation for ellipticals. Their large-aperture absolute
magnitude is $M_V(z=1.55\mid \Omega=1) \simeq -21.62 -5 \log_{10} h$ (measured
direct in the rest frame). This yields to $\sigma_v= 222 \; {\rm to}\; 292 \;
{\;{\rm km\,s^{-1}}}$, which is a very reasonable range for a giant
elliptical, and is adopted in the following analysis.

Having established an abundance and an equivalent circular velocity for these
galaxies, the treatment of them will differ in one critical way from the
Lyman-$\alpha$ and Lyman-limit galaxies.  For these, the normal
Press-Schechter approach assumes the systems under study to be newly born. For
the Lyman-$\alpha$ and Lyman-limit galaxies, this may not be a bad
approximation, since they are evolving rapidly and/or display high levels of
star-formation activity. For the radio galaxies, conversely, their inactivity
suggests that they may have existed as discrete systems at redshifts much
higher than $z\simeq 1.5$.  The strategy will therefore be to apply the
Press-Schechter machinery at some unknown formation redshift, and see what
range of redshift gives a consistent degree of inhomogeneity.

\begin{figure}
\vspace*{-1.0cm}
\hspace*{-0.5cm}
\psfig{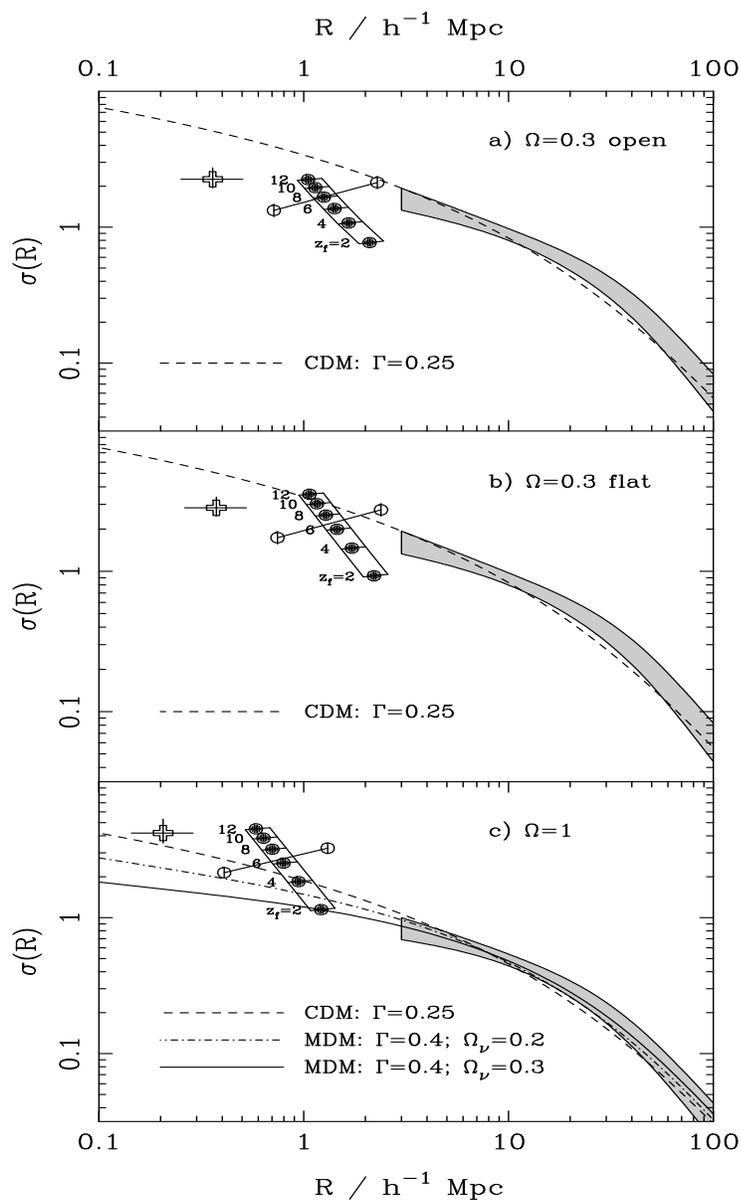}
\caption{The present-day linear fractional rms fluctuation in density averaged
in spheres of radius $R$. The data points are Lyman-$\alpha$ galaxies (open
cross) and Lyman-limit galaxies (open circles) The diagonal band with solid
points shows red radio galaxies with assumed collapse redshifts 2, 4, \dots
12.  The vertical error bars show the effect of a change in abundance by a
factor 2.  The horizontal errors correspond to different choices for the
circular velocities of the dark-matter haloes that host the galaxies.  The
shaded region at large $R$ gives the results inferred from galaxy clustering.
The lines show CDM and MDM predictions, with a large-scale normalization of
$\sigma_8=0.55$ for $\Omega=1$ or $\sigma_8=1$ for the low-density models.}
\end{figure}

\subsection{Collapse redshifts and ages for red radio galaxies}

Fig.~1 shows the $\sigma(R)$ data which result from the Press-Schechter
analysis, for three cosmologies. The $\sigma(R)$ numbers measured at various
high redshifts have been translated to $z=0$ using the appropriate linear
growth law for density perturbations.

The open symbols give the results for the Lyman-limit (largest $R$) and
Lyman-$\alpha$ (smallest $R$) systems. The approximately horizontal error bars
show the effect of the quoted range of velocity dispersions for a fixed
abundance; the vertical errors show the effect of changing the abundance by a
factor 2 at fixed velocity dispersion.  The locus implied by the red radio
galaxies sits in between. The different points show the effects of varying
collapse redshift: $z_c=2, 4, \dots, 12$ [lowest redshift gives lowest
$\sigma(R)$]. Clearly, collapse redshifts of 6 -- 8 are favoured for
consistency with the other data on high-redshift galaxies, {\it independent of
theoretical preconceptions and independent of the age of these galaxies}(see
also Kashlinsky \& Jimenez (1997)).

What is then the age of the red radio galaxies as inferred by their high
collapse redshifts? First bear in mind that in a hierarchy some of the stars
in a galaxy will inevitably form in sub-units before the epoch of collapse.
At the time of final collapse, the typical stellar age will be some fraction
$\alpha$ of the age of the universe at that time: ${\rm age} = t(z_{\rm obs})
- t(z_c) + \alpha t(z_c)$.  We can rule out $\alpha=1$ (i.e. all stars forming
in small subunits just after the big bang). For present-day ellipticals, the
tight colour-magnitude relation only allows an approximate doubling of the
mass through mergers since the termination of star formation (Bower at
al. 1992). This corresponds to $\alpha\simeq 0.3$ (Peacock 1991). A non-zero
$\alpha$ just corresponds to scaling the collapse redshift as ${\rm apparent}\
(1+z_c)\propto (1-\alpha)^{-2/3}$, since $t\propto (1+z)^{-3/2}$ at high
redshifts for all cosmologies.  For example, a galaxy which collapsed at $z=6$
would have an apparent age corresponding to a collapse redshift of 7.9 for
$\alpha=0.3$.

Converting the ages for the galaxies to an apparent collapse redshift depends
on the cosmological model, but particularly on $H_0$.  Some of this
uncertainty may be circumvented by fixing the age of the universe. After all,
it is of no interest to ask about formation redshifts in a model with
e.g. $\Omega=1$, $h=0.7$ when the whole universe then has an age of only 9.5
Gyr. If $\Omega=1$ is to be tenable then either $h<0.5$ against all the
evidence or there must be an error in the stellar evolution timescale. If the
stellar timescales are wrong by a fixed factor, then these two possibilities
are degenerate. It therefore makes sense to measure galaxy ages only in units
of the age of the universe -- or, equivalently, to choose freely an apparent
Hubble constant which gives the universe an age comparable to that inferred
for globular clusters.  In this spirit, Fig. 2 gives apparent ages as a
function of effective collapse redshift for models in which the age of the
universe is forced to be 14 Gyr (e.g. Jimenez et al. 1996, see also Fusi-Pecci
in this volume).

\begin{figure}
\vspace*{-3.0cm}
\hspace*{2.0cm}
\psfig{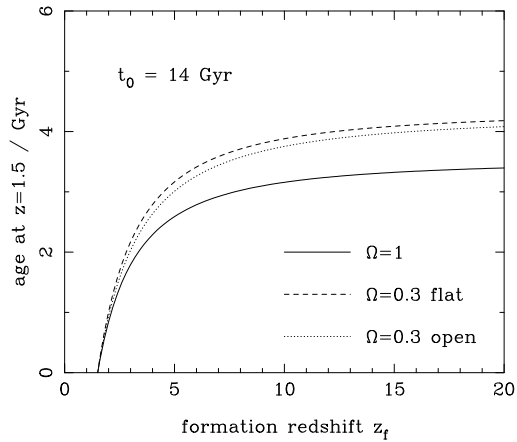}
\vspace*{-3.0cm}
\caption{Age of the Universe at $z=1.5$ for several cosmologies, it 
transpires from the figure that the age of galaxies formed at $z>5$ is well
constrained between 3 and 4 Gyr}
\end{figure}

This plot shows that {\it the ages of the red radio galaxies are not permitted
very much freedom}. Formation redshifts in the range 6 to 8 predict an age of
close to 3.1 Gyr for $\Omega=1$, or 3.7 Gyr for low-density models,
irrespective of whether $\Lambda$ is nonzero.  The age-$z_c$ relation is
rather flat, and this gives a robust estimate of age once we have some idea of
$z_c$ through the abundance arguments.

\section{The age of 53w069}

As seen from the previous section, galaxies at $z \sim 1.5$ with low
comoving densities, are excellent sites for setting the stellar clock since
their age is known from other independent arguments. In what remains I will
analyse how some of the different set of synthetic stellar population models
available perform when trying to recover the age of 53W069 (the reddest
passively evolving galaxy found at high$-z$).

\begin{figure}
\vspace*{-5.0cm}
\hspace*{0.7cm}
\centerline{
\psfig{figure=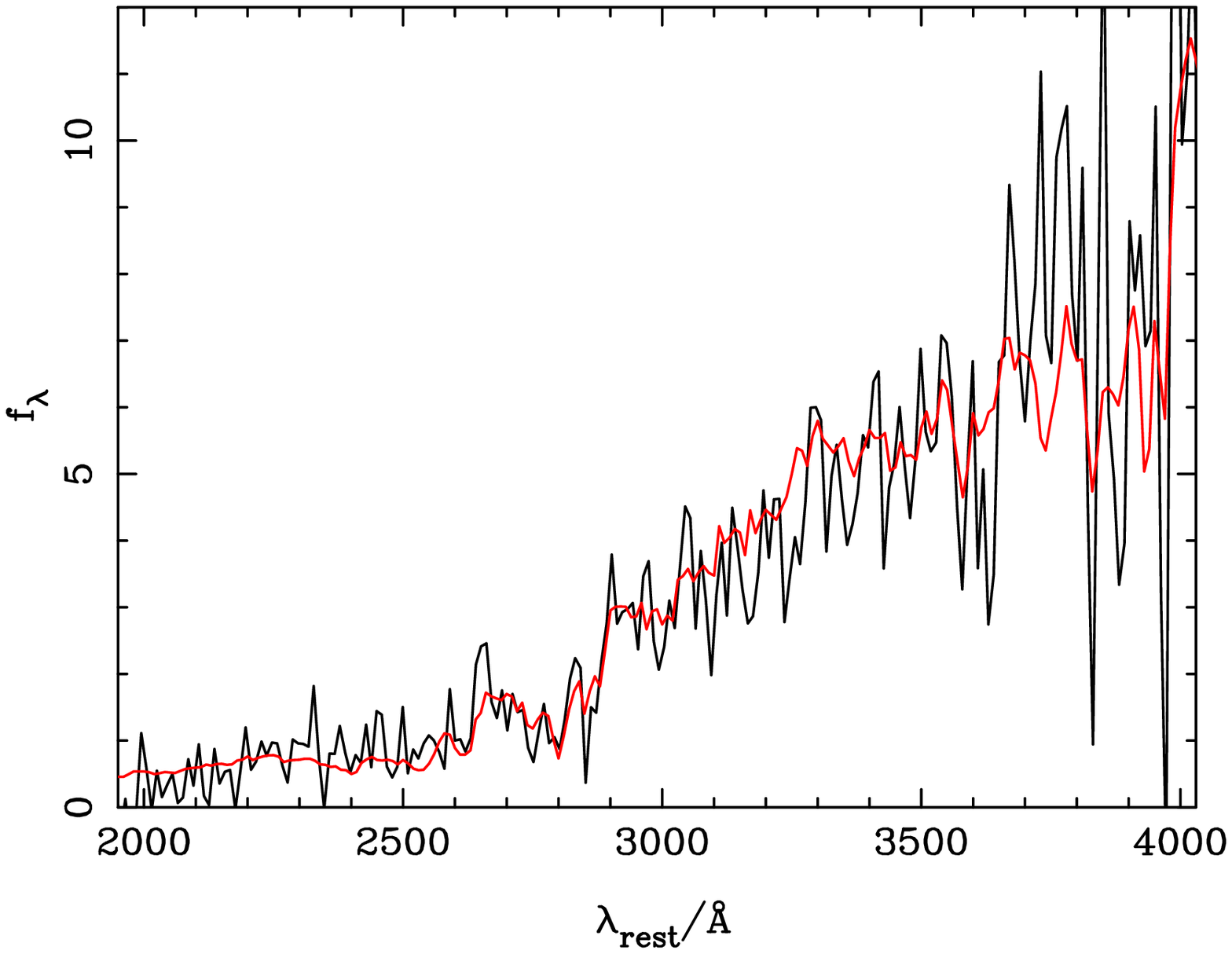,height=11cm,angle=0}
\hspace*{-2.0cm}
\psfig{figure=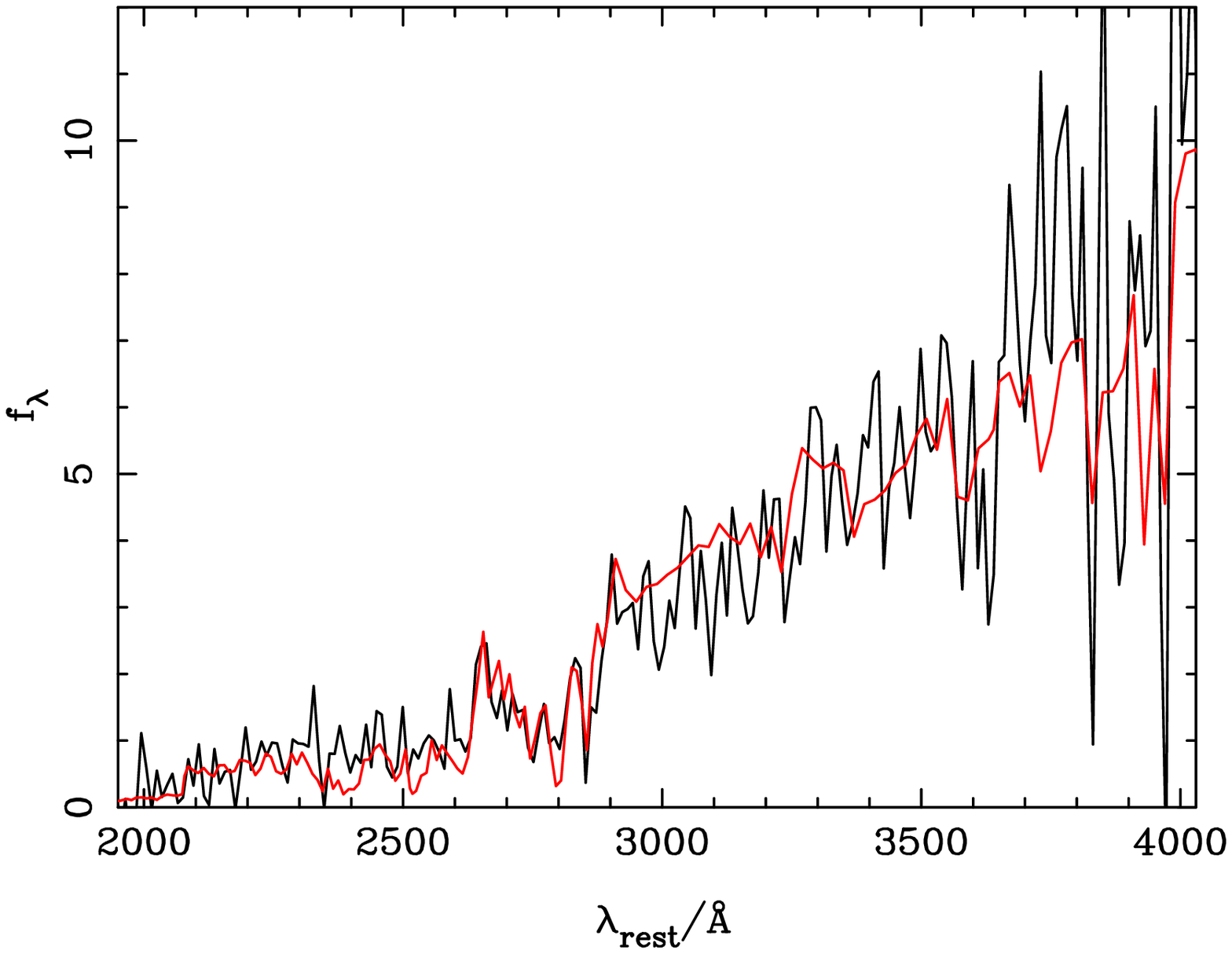,height=11cm,angle=0}}
\caption{Left panel: the spectrum of 53W069 overlaid with the best fitting
Bruzual-Charlot model, which has an age of 3.3 Gyr. Right panel: Same as
before but now overlaid with the best fitting model from Jimenez et
al. (1999b), which has an age of 4.0 Gyr.}
\end{figure}

The spectral energy distribution (SED) of 53W069 is presented in Fig.~3. In
order to determine its age I have performed properly weighted chi-squared fits
to the ultra-violet SED using the popular Worthey and also Bruzual \& Charlot
synthetic stellar population models and the models developed by our group
(Jimenez et al. 1999b). The results are listed in Table~1. A few important
points transpire from this table. First, all models yield ages larger than 3
Gyr for 53W069. Second, column 1 and 2 show that some models seem to be
internally inconsistent, in the sense that they are capable of reproducing
very red $R-K$ colours at a much younger age than they can reproduce the
ultraviolet SED or the spectral breaks. However, since $R-K$ is mainly
affected by the evolution of the late stages of stellar evolution, once should
focus on ages derived from the UV-spectrum since this only depends on the
correct prediction of the MSTO, a simpler and thus an easier part to
model. Indeed, not do so it tantamount to throwing away the new, more robust
information which can be gleaned from the spectroscopy. If one focuses on the
results of 53W069 then, ignoring the anomalously young $R-K$ age produced by
some models, all sets are basically in good agreement that the overall shape
of the UV SED is consistent with an age in the range 3.0 to 4.0 Gyr, and
certainly yielding a robust (99\% confidence) minimum age of 3.0 Gyr. Thus one
finds that {\it all} models yield basically to the same result, in perfect
agreement with the independently derived age in section 2. Therefore, one can
conclude that the stellar clock is properly set, even at these young ages. The
anoumalously small ages (about 1 Gyr) obtained by fitting the $R-K$ colours,
can be entirely attributed to the fact that $K$ (rest-frame 1 $\mu$) is
entirely dominated by the giants population, and thus by mass loss and other
complications from stellar evolution during its late stages.

Spectroscopically, 53W069 thus appears to be the best known example of old,
passively-evolving elliptical galaxies at redshifts as high as $z \sim
1.5$. It is worth noting that $z \sim 1.5$ is the redshift where one expects 
galaxies to be redder, and not at higher redshift (see Jimenez et al. 1999a)

\begin{table}
\begin{center}
\footnotesize\rm
\caption{A comparison of age estimates for the stellar population of 53W069 as
derived from the instantaneous burst models of Bruzual \& Charlot (B\&C),
Worthey (1998) (W) and Jimenez et al (1999b) (J99), when used to fit different
spectral indicators of age.}
\vspace*{.5cm}
\begin{tabular}{llll}
\hline
{\bf Feature} & {\bf B\&C} & {\bf W98} & {\bf J99} \\
\hline
UV-SED   & 3.3 Gyr & 3.1 Gyr & 3.7 Gyr \\
$R-K$    & 1.6 Gyr & 2.0 Gyr & 4.0 Gyr \\
2649 \AA & 5.0 Gyr & 4.0 Gyr & 4.0 Gyr \\
2900 \AA & 4.5 Gyr & 4.2 Gyr & 4.0 Gyr \\
\hline
\end{tabular}
\end{center}
\end{table}

In summary, the new data on 53W069 clearly show that the Universe at $z \sim
1.5$ contains stellar systems whose populations are 3 to 4 Gyr old. At $z \sim
1.5$ the Universe was less than 30\% of its present age, and the uncertainties
are largely independent of those encountered in GCs studies. The existence of
53W069 permits only low Hubble constants and/or low cosmic densities; in
particular, an $\Omega=1$ Universe requires $H_0 \leq 45$ km s$^{-1}$
Mpc$^{-1}$ . On the other hand, in an Universe with a cosmological constant,
53W069 requires $\Omega_{\lambda} \geq 0.4$ if $H_0 \geq 60$ km s$^{-1}$
Mpc$^{-1}$

\acknowledgments

This paper draws on unpublished collaborative work with James Dunlop, John
Peacock, Arjun Dey, Hyron Spinrad, Dan Stern and Rogier Windhorst.


\begin{references}
\reference Chambers, K.C., Charlot, S., 1990, \apj, 348, L1
\reference Dey et al., 1999, in preparation
\reference Dickinson M., 1995, in ``Fresh Views of Elliptical Galaxies'', ASP conf. ser. Vol 86,  eds A. Buzzoni, A. Renzini, A. Serrano, p283
\reference Dunlop, J., 1998, astro-ph/9801114
\reference Dunlop J., Peacock J., Spinrad H., Dey A., Jimenez R., Stern D.,
Windhorst R., 1996, Nature, 381, 581
\reference Efstathiou G., Rees M.J., 1988, {MNRAS}, {230}, 5P
\reference Eke V.R., Cole S., Frenk C.S., {1996}, {MNRAS}, {282}, {263}
\reference Jimenez R., Thejll P., J\o rgensen U.G., MacDonald J., Pagel B., 
1996, {MNRAS}, 282, 926
\reference Jimenez et al., 1999a, MNRAS, 305, L16
\reference Jimenez et al., 1999b, MNRAS, in press
\reference Kashlinsky A., Jimenez R., {\apj}, {474}, L81
\reference Ma C., Bertschinger E., 1994, {\apj}, {434}, L5
\reference Peacock J.A., 1991, in ``Physical Cosmology'', proc. 2$^{nd}$
Rencontre de Blois, eds A. Blanchard, L. Celnekier, M. Lachi\`eze-Rey \& J. 
Tr\^an Thanh V\^an (Editions Fronti\`eres), p337
\reference Peacock J.A., Jimenez R., Dunlop J.S., Waddington I., Spinrad H.,
Stern D., Dey A., Windhorst R.A., 1998, {MNRAS}, 296, 1089
\reference Press W.H., Schechter P., 1974, {\apj}, {187}, 425
\reference Spinrad H., Dey A., Stern D., Dunlop J., Peacock J., Jimenez R., Windhorst R., 1997, \apj, 484, 581
\reference Steidel et al., 1998, astro-ph/9811399
\reference White S.D.M., Efstathiou G., Frenk C.S., 1993, {MNRAS}, {262}, 1023
\end{references}
\end{document}